\documentclass[
12pt,
preprint,
 amsmath,amssymb,
]{iopart}
\usepackage{graphicx}% Include figure files
\usepackage{dcolumn}% Align table columns on decimal point
\usepackage{bm}% bold math
\usepackage{upgreek}
\usepackage{hyperref}% add hypertext capabilities
\usepackage[mathlines]{lineno}% Enable numbering of text and display math
\usepackage{url}
\usepackage{setspace}
\usepackage{color}
\usepackage{changes}
\date{\today}% It is always \today, today,
\begin{document}
\title[Probing hotspots of PERS by nanomanipulation of CNTs]{Probing hotspots of plasmon-enhanced Raman scattering by nanomanipulation of carbon nanotubes}
\author{Sebastian Heeg\footnote{Present address:
Photonics Laboratory, ETH Z\"urich, 8093 Z\"urich, Switzerland}, Nick Clark, Aravind Vijayaraghavan}
\address{School of Materials, The University of Manchester, Manchester M13 9PL, UK}
\ead{sheeg@ethz.ch}
\begin{abstract}
We present a two-step procedure to probe hotspots of plasmon-enhanced Raman scattering with carbon nanotubes. Dielectrophoretic deposition places a small carbon nanotube bundle on top of plasmonic Au nanodimer. After 'pre-characterising' both the nanotubes and dimer structure, we subsequently use the tip of an AFM to push the bundle into the plasmonic hotspot located in the $25\,$nm wide dimer gap, characterize its location inside the gap, and observe the onset of plasmon-enhanced Raman scattering. Evidence for the activation of the carbon nanotube's double-resonant D-mode by the near-field of the plasmonic hotspot is discussed. 
\end{abstract}
\submitto{Nanotechnology}
\maketitle

\newpage
%INTRO1: EM & CENH
%\textbf{Keywords: plasmon-enhanced Raman scattering, carbon nanotubes, plasmonic hotspot, nano-manipulation}
%\newline
%\newline
\section{Introduction}
Metallic nanostructures enable the localization of electromagnetic waves into nanoscale volumes far smaller than the wavelength of light at optical frequencies~\cite{Maier:2007wq,Novotny:2012tia}. The electromagnetic wave couples to the collective excitation of electrons in the metal nanostructure called localized surface plasmon resonance (LSPR). This leads to a strong enhancement of the local near-field at the metal surface. When two nanostructures are closely spaced, coupling of the confined LSPRs (at certain polarizations) leads to a collective excitation across both, generating particularly strong field enhancement in the gap between them. Such a region is termed plasmonic hotspot. The field enhancement varies considerably inside and around such a plasmonic hotspot and may differ by several order of magnitude over tens of nanometers~\cite{LeRu:2012fp,Maier:2007wq,Novotny:2012tia,Gramotnev:2010jk}.

One of the most spectacular applications of LSPRs is plasmon-enhanced Raman scattering (PERS), where the incident light and that inelastically scattered from an object inside a plasmonic hotspot is greatly enhanced~\cite{Fleischmann:1974jh,Willets:2007ko}. PERS has enabled the detection of single molecules and offers great potential for studying the fundamental aspects of light-matter interaction as well as biological and chemical sensing applications.~\cite{Kneipp:1997ua,Nie:1997dl,ISI:000241099100013,Roelli:2015bv,Mueller:2016kya,Schmidt:2016ek,Jorio:2017bl,2016Sci...354..726B,Sharma:2012vp,Halas:2011cz,Stockmann:2015ie}

A recurring problem in quantifying PERS are small geometric variations between plasmonic structures of the same design, which may strongly alter the enhancement experienced by a Raman scatterer that is nominally located at the same position inside a corresponding plasmonic hotspot~\cite{LeRu:2009id,Darby:2015kn}. It is therefore desirable to probe the PERS enhancement at different locations in and around one plasmonic hotspot using the same Raman scatterer. \textit{Kusch et al.} recently suggested to read out the PERS enhancement at a plasmonic hotspot via the Raman signal of a sharp Si-tip of a scanning near-field optical microscope~\cite{Kusch:2017de}. The spatial extension of the tip, however, masks subtle variations of the enhancement on the nanometer scale. Similar restrictions apply to molecules, the traditional probe for PERS, because their exact location and orientation inside a plasmonic hotspot are impossible to control and, more importantly, cannot be altered. A nanoscale Raman scatterer that probes the enhancement in and around a plasmonic hotspot is therefore highly desirable.  

\begin{figure}[t]%
\centering
\includegraphics*[width=8.5cm]{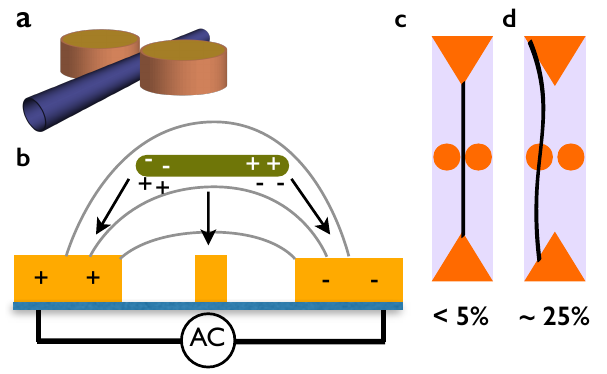}
\caption{(a) Schematic of CNT at the plasmonic hotspot in a nanodimer gap. (b) AC voltage drives the CNTs in solution to deposit between the electrodes, where the plasmonic dimers are located. Based on Ref.~\cite{Heeg:2014cn}, DEP has a $<5\%$ yield for nanotubes passing through the gap, as in (c), and $\sim25\%$ for nanotubes crossing one of the discs forming the dimer in (d).}
\label{FIG:DEP_YIELD}
\end{figure}

Carbon nanotubes (CNTs) overcome some of these limitations and have recently emerged as a promising alternative to probe and quantify plasmon-enhanced Raman scattering on the nanoscale as their one-dimensional nature allows us to experimentally obtain the exact location and orientation within a plasmonic hotspot~\cite{Assmus:2007fw,Scolari:2008kt,Cancado:2009vu,Heeg:2014cn,Heeg:2014kx,Bauml:2017td,Paradiso:2015jx,Mueller:2017hb}. Interfacing CNTs with rationally designed hotspots, on the other hand, remains a challenge. We recently demonstrated the benefit of using directed dielectrophoretic assembly (DEP) to place carbon nanotubes precisely into the plasmonic hotspot at the $\sim25\,$nm gap of gold nanodimers, Fig.~\ref{FIG:DEP_YIELD}(a), and observed a $10^3-10^4$ enhancement of the Raman intensity~\cite{Heeg:2014cn,Heeg:2014kx}. More importantly, CNTs are perhaps the only nanoscale object that is able to probe PERS enhancement at different locations in a plasmonic hotspot, as the location and orientation of a nanotube can be purposefully alterted by nanomanipulation.~\cite{Falvo1997fa,Hertel:1998gs}

Here we demonstrate a two-step scheme that combines dielectrophoretic deposition of carbon nanotubes with tip-based nanomanipulation to probe plasmon-enhanced Raman scattering outside and inside the plasmonic hotspot of a gold nanodisc dimer. With the tip of an atomic force microscope (AFM) we push a small carbon nanotube bundle, placed on one of the discs forming the dimer by dielectrophoresis, into the dimer gap. We determine the position of the CNT bundle inside the hotspot and verify the onset of plasmon-enhanced Raman scattering by an $100$-fold Raman enhancement of the nanotube's G-mode, an inverted polarization behaviour and the spatial localization of the Raman signal after nano-manipulation. Beyond quantifying plasmon-enhanced Raman scattering, the proposed scheme will enable the realization of nanotube-nanoplasmonic experimental systems that were previously not accessable.

%Fig
%Fig

\section{Experimental}
The sample fabrication and dielectrophoretic deposition process follows the procedures described in Refs.~\cite{Heeg:2014cn} and~\cite{Heeg:2014kx}. In short, arrays of dimers and electrodes for DEP were fabricated on a SiO$_2$ ($90\,$nm) on Si substrate by electron beam lithgraphy using a LEO 1530 Gemini FEG SEM and a Raith Elphy Plus Lithography System. Metallization was carried out by evaporating $5\,$nm Cr + $40\,$ nm Au followed by lift-off in an ultrasonic bath. The dimers are located between electrode pairs (distance $\sim1\,\upmu$m) with sharp tips for directed dielectrophoretic deposition. During directed dielectrophoretic deposition, a droplet of ultrapure, unsorted CNTs (www.nanointegris.com) in aqueous solution is placed on top of the arrays and a kHz AC voltage is applied between the electrodes~\cite{Krupke:2003cl,Vijayaraghavan:2009gv}. Initial AFM characterization after DEP was performed using an Park Systems XE 150 AFM. 
Nanomechanical manipulation was performed using a Bruker Dimension Icon AFM using Bruker TAP150A probes with cantilevers with a nominal stiffness of $5\,$N/m. All imaging was performed in non-contact mode. Raman characterization was performed using Horiba Yobin Ivon XploRa ($\lambda=532\,$nm) and Witec alpha300 ($\lambda=633\,$nm) single grating spectrometers using $100\times$ objectives (NA $0.9$), and piezo stages for spatial mapping. To measure the polarization, we rotated the sample by $90^{\circ}$. We used integration times up to $10\,$s and laser powers between $1\,$mW ($532\,$nm) and $50\,\upmu$W ($633\,$nm). The luminescence background from the Au nanostructures was subtracted for all Raman spectra.

\section{Results and Discussion}
We fabricated arrays of dimer structures consisting of two closely spaced nanodiscs by electron beam lithography on top of a SiO$_2$/Si substrate (see Methods). The dimer's dimensions (cylinders with diameter $\sim100\,$nm, height $45\,$nm, separated by a gap of $\sim25\,$nm) were chosen to provide resonant enhancement for $\lambda=633\,$nm excitation polarized along the dimer axis~\cite{Heeg:2013di,Heeg:2014cn,Wasserroth:2018bm}. The strongest coupling to the high-intensity near-field occurs for a nanotube placed inside the plasmonic hotspot at the nanoscale gap of the dimer~\cite{Maier:2007wq,Novotny:2012tia}. We realized such an interface for plasmon-enhanced Raman scattering of CNTs using directed dielectrophoretic deposition from solution as described in detail in Refs.~\cite{Heeg:2014cn} and \cite{Heeg:2014kx}. Dielectrophoretic forces drive the nanotube to deposit between electrodes where we have placed the plasmonic dimers, c.f. Fig.~\ref{FIG:DEP_YIELD}(b). Ideally the CNT connects the electrode tips in a straight line and passes through the gap as shown in Fig.~\ref{FIG:DEP_YIELD}(c). This ideal configuration, however, has a yield below $5\%$. A scenario where the nanotube does not connect the electrode tips directly and crosses one of the discs forming the dimer as shown Fig.~\ref{FIG:DEP_YIELD}(d) is much more likely ($\sim25\%$) but does not provide significant Raman enhancement. It is therefore the ideal starting point for the two step procedure proposed in this work, because it allows us to characterize the CNT and its Raman signatures before it is interfaced with the plasmonic hotspot at the dimer gap. 

%FIG2
\begin{figure}[pt]%
\centering
\includegraphics*[width=8.5cm]{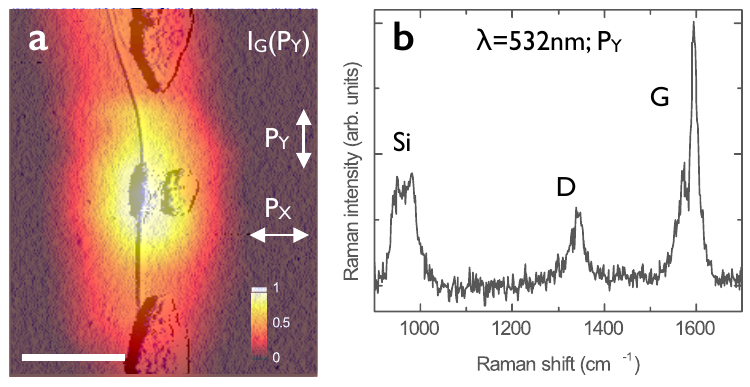}
\caption{(a) AFM phase image of CNT-B overlaid with its integrated G-peak intensity for $532\,$nm and P$_Y$. The scale bar is $500\,$nm and the intensity normalized to the maximum intensity. (b) Raman spectrum acquired at the dimer. D-mode and G-mode of CNT-B and the second order Si peak are labelled in the spectrum.}
\label{FIG:CNT-B_532}
\end{figure}
%END FIG2

We show an AFM phase image of such an initial scenario in Fig.~\ref{FIG:CNT-B_532}(a). A small CNT-bundle (height $\sim 7\,$nm, labelled CNT-B) crosses the left disc of a plasmonic dimer. The phase image is overlaid with a spatial map of the nanotube G-band just below $1600\,$cm$^{-1}$, the most prominent Raman feature in CNTs~\cite{Thomsen:2007vc}. The corresponding Raman spectrum, taken with the laser centered on the dimer, is shown in Fig.~\ref{FIG:CNT-B_532}(b). The D-mode at $\sim1350\,$cm$^{-1}$ indicates the presence of defects. It also appears in Raman spectra of CNTs dropcasted from the as purchased solution on a bare SiO$_2$/Si substrate and is therefore not caused by DEP~\cite{Heeg:2014cn,Heeg:2014kx}. Both the energy of the excitation ($\lambda=532\,$nm) and its polarization P$_Y$ (perpendicular to the dimer) prevent plasmonic enhancement from the dominant dipolar LSPR of the dimer. 

The spatial distribution of the integrated G-mode intensity $I_G(P_Y)$ in Fig.~\ref{FIG:CNT-B_532}(a) matches that of CNT-B.  We attribute the increased intensity around the dimer disc to a maximized overlap between CNT-B and the laser spot in combination with minor plasmonic enhancement from the disc's LSPR or due to electric field line crowding at the disc edges~\cite{Maier:2007wq}. We did not observe a considerable signal for P$_X$, as the strong antenna effect in CNTs suppresses Raman scattering for polarizations perpendicular to the tube axis~\cite{Thomsen:2007vc,Reich:2009wo}. For an excitation wavelength that overlaps with the dimer resonance ($\lambda=633\,$nm), no Raman signal within our detection threshold was observed for both P$_X$ and P$_Y$. 

%FIG pushit
\begin{figure}[hpt]%
\centering
\includegraphics*[width=8.5cm]{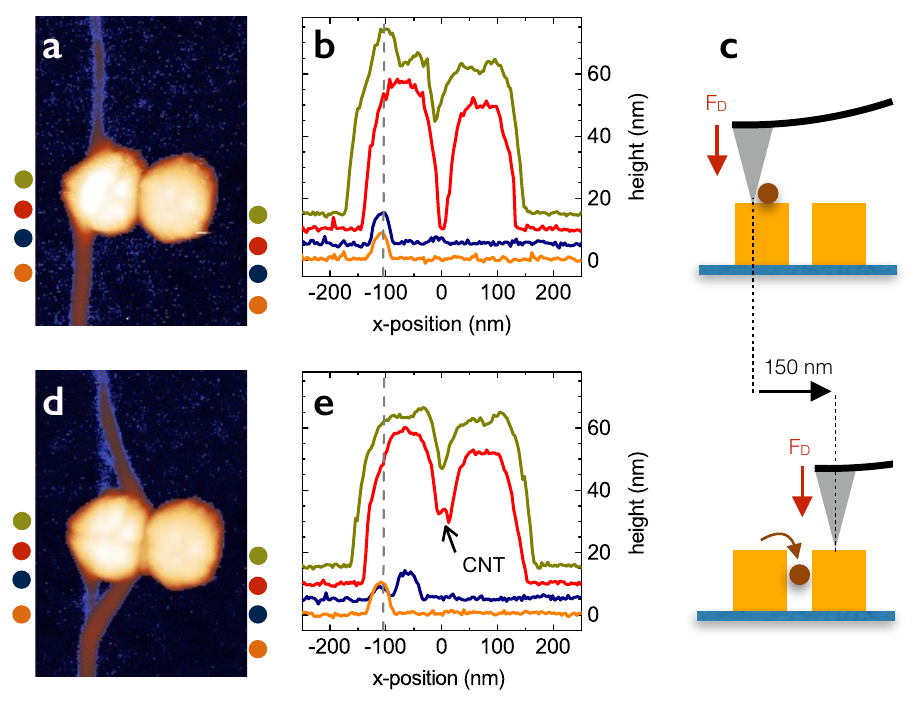}
\caption{(a) AFM topography of CNT-B crossing the left dimer disc. (b) AFM height profiles connecting the dots of the corresponding colour in (a). The grey dashed line indicates the lateral x-position of CNT-B on top and next to the dimer dics. (c) The downward force F$_D$ holds down the tip while it is moved across the dimer, thereby moving CNT-B into the plasmonic hotspot at the gap. (d) AFM topography after nanomanipulation. (e) Height profiles confirming that CNT is located in the gap.}
\label{FIG:PUSHIT}
\end{figure}

Before we discuss the nano-manipulation of CNT-B, we investigate more closely the topography of CNT-B and its interface with the dimer structure. A high-resolution AFM topography image is shown in Fig.~\ref{FIG:PUSHIT}(a). It confirms that CNT-B crosses the left nanodisc. Height profiles parallel to the dimer axis are shown in Fig.~\ref{FIG:PUSHIT}(b). They connect the dots of corresponding colour in Fig.~\ref{FIG:PUSHIT}(a), and are offset by $5\,$nm with respect to each other for clarity. On top of the dimer (green), CNT-B sits close to the edge of the left disc as indicated by the dashed vertical line. The AFM tip is insufficiently sharp to probe between the dimers. The tip sides as depicted in Fig.~\ref{FIG:PUSHIT}(b) are angled at $15^{\circ}$ deg  (left) and $25^{\circ}$ (right) to the vertical. This leads to a slight asymmetry in the observed height profile at the sides of the discs. Close to the dimer edge (red), the CNT-B adheres to the disc wall and does not show a topographic feature. The tip, on the other hand, reaches to the bottom of the dimer gap. At distances of $20\,$nm (blue) and $100\,$nm (orange) away from the dimer, CNT-B adheres to the substrate at the same lateral x-position. 

To move CNT-B into the gap, we ramped the tip into a chosen point on the dimer disc - to the left of CNT-B/dashed lines in Fig.~\ref{FIG:PUSHIT}(b,c) - until a deflection setpoint of $50\,$nm was reached. This value corresponds to a downward force $F_D$ of approximately $200-300\,$nN. While maintaining the downward force, the tip was slowly ($50\,$nm/s) moved laterally $~150\,$nm across the dimer gap to a second chosen point on the right nanodiscs, bottom Fig.~\ref{FIG:PUSHIT}(c), and retracted. The tip was then immediately used in tapping mode to image the dimer and CNT, Fig.~\ref{FIG:PUSHIT}(d). We achieved the same resolution as in Fig.~\ref{FIG:PUSHIT}(a) prior to the nanomanipulation -- the observed profile of CNTB is the real profile convoluted with the tip apex profile -- which shows that the tip did not measureably blunt while in contact with the surface. 

The topography in Fig.~\ref{FIG:PUSHIT}(d) reveals that CNT-B was moved to the right such that it passes through the dimer gap. We will now deduce the exact configuration of the nanotube in the gap by comparing the corresponding height profiles in Fig.~\ref{FIG:PUSHIT}(e) to those take before moving CNT-B in Fig.~\ref{FIG:PUSHIT}(b). On top of the dimer (green), CNT-B has been removed from previous location (dashed line). At the dimer edge (red), CNT-B appears as a feature in the gap at a height of $\sim25\,$nm, preventing the tip from imaging the gap region below. Around $20\,$nm away from the dimer (blue), CNT-B has been shifted to the right to accommodate for its new location at the dimer gap without building up strain. Further away from the dimer (gap), no changes in the tube's position have occured. 

The observed topography before and after nanomanipulation indicates that CNT-B is moved to a position suspended at a height of $25\,$nm in the gap. At first, the tip pushes CNT-B across the left disc's surface, thereby overcoming adhesion of the CNT-B on the sides of the disc and on the substrate. The tip then enters the dimer gap to maintain F$_D$ and carries the tube in the gap. The tip then retracts from the gap while the nanotube slips off and remains inside the gap. The position at CNT-B at a height of around $25\,$nm is in good agreement with the penetration depth of our tip with $25^{\circ}$ side angle and a gap width of $25\,$nm. Note that during imaging in tapping mode, along the dimer axis the tip only reaches $10\,$nm into the gap, Fig.~\ref{FIG:PUSHIT}(e), and is therefore unable to probe CNT-B. This limitation also applies for mapping out the enhancement arising from a plasmonic dimer through the Raman signal of a Si-tip mounted on a scanning probe microscope as recently suggested by \textit{Kusch et al.}~\cite{Kusch:2017de}.

%FIG4
\begin{figure}[pt]%
\centering
\includegraphics*[width=8.5cm]{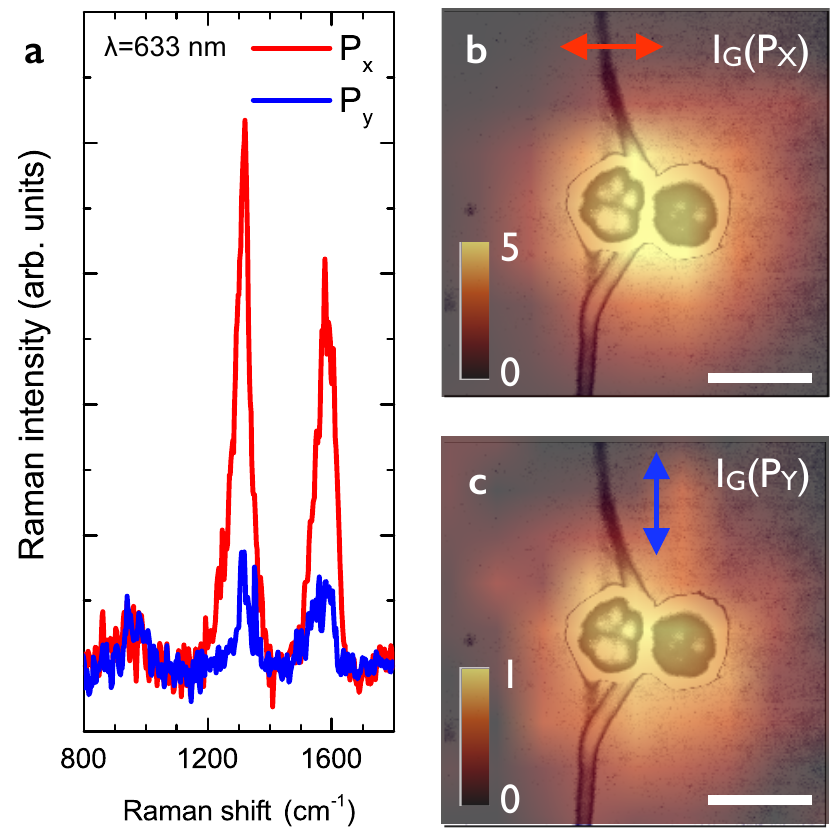}
\caption{(a) PERS spectra from CNT-B after nanomanipulation for P$_X$ (red) and P$_Y$ (blue). Au photoluminescence has been substracted from the spectra. (b) and (c) show Raman maps of the integrated G-mode intensity $I_G$ for P$_X$ and P$_Y$, respectively. They are overlayed with the AFM topography from Fig.~\ref{FIG:PUSHIT}(d). The scale bar in (b) and (c) is $200\,$nm.}
\label{FIG:RAMAN_ENH}
\end{figure}
The observation of plasmon-enhanced Raman scattering from CNT-B after nanomanipulation confirms that we have successfully interfaced CNT-B with the plasmonic hotspot in the dimer gap. Figure~\ref{FIG:RAMAN_ENH}(a) shows Raman spectra of CNT-B measured on the dimer with $\lambda=633\,$nm for both P$_X$ (red) and P$_Y$ (blue). The Raman intensity for P$_X$ dominates the spectrum, and is a clear sign of enhancement due to near-field from the dipolar LSPR of the dimer structure. The signal is much stronger than for the non-resonant case ($\lambda=532\,$nm), where we did not observe any signal for P$_X$. It is also stronger than for P$_Y$, along the CNT-axis, compare Fig.~\ref{FIG:CNT-B_532}(b). We have previously identified this inverted polarization behaviour as a clear sign of plasmonic enhancement from CNTs in dimer hotspots~\cite{Heeg:2014cn,Heeg:2014kx}. 

The spatial distribution of the G-mode signal for P$_X$ is shown in Fig.~\ref{FIG:RAMAN_ENH}(b). It is strongly localized in space and shows the point-like character of the Raman enhancement arising from hotspot the dimer gap. Compared to P$_X$, the relative magnitude of the enhanced signal for P$_Y$ and its localization, Fig.~\ref{FIG:RAMAN_ENH}(c), may seem surprising. It originates from minor plasmonic enhancement that couples favourably to CNT-B because the near field is polarized parallel to the axes of the tubes forming CNT-B. This is in stark contrast to the Raman intensity for P$_X$ that scales only with the projection of the near field polarization on the tube axis~\cite{Heeg:2013di,Heeg:2014cn}, explaining the relative intensities in Fig.~\ref{FIG:RAMAN_ENH}(c). From the inverted polarization behaviour and signal localization presented in Figs.~\ref{FIG:RAMAN_ENH}(b) and (c), we estimate $2.7\times10^2$ as the lower limit of the enhancement factor of the G-mode, details see Ref.~\cite{Heeg:2014cn}. 

The D-mode is the dominant feature in our PERS spectra, Fig.~\ref{FIG:RAMAN_ENH}(b). The ratio $I_D/I_G$ is typically regarded as a measure for the defect concentration in nanoscale graphitic material~\cite{Thomsen:2007vc}. For CNT-B, it increased from $0.43$ ($532\,$nm, P$_Y$) before the nanomanipulation to $0.61$ ($633\,$nm, P$_Y$) and $1.22$ ($633\,$nm, P$_X$) in the presence of enhancement. For bulk quantities of our CNT starting material and for small CNTs bundles deposited directly into the dimer gap by DEP, the $I_D/I_G$ ratio increases upon changing the excitation wavelength from $532\,$nm to $633\,$nm, see Ref.~\cite{Heeg:2014cn}. Therefore, the increase in $I_D/I_G$ for P$_Y$ is intrinsic to the CNTs used in this work and not caused by nanomanipulation. Did moving CNT-B, however, induce structural defects at tube segments now located at the plasmonic hotspot, which could explain the increase in $I_D/I_G$ from $0.61$ ($633\,$nm, P$_Y$) to $1.22$ ($633\,$nm, P$_X$)? Previous studies showed that moving CNTs by AFM does not damage the nanotubes~\cite{Duan:2007gv,Mussnich:2015iq}. \textit{Yano} and co-workers moved carbon nanotubes with an AFM tip and subsequently characterized them by tip-enhanced Raman spectroscopy~\cite{Yano:2013ft}. They did not observe any defects at the manipulated tube segments while probing with a spatial resolution of $20\,$nm, comparable to the localized nature of the enhancement in this work. As we were moving a bundle of CNTs, any stretching of the  bundle due to nanomanipulation can be relaxed by interfacial sliding between tubes in the bundle which occurs at a much lower force than the introduction of structural defects within the nanotube~\cite{Zhang:2012ht}. We further exclude radiation-induced damage because we used low powers ($50\,\mu$W) when acquiring the PERS spectra at $\lambda=633\,$nm.

In the light of these studies, we argue that the increase in the relative D-mode intensity in our PERS spectra is not caused by structural defects. Instead, it is a consequence of the strongly localized nature of the near-field in the plasmonic hotspot. A recent study observed a strong D-mode for defect-free graphene interfaced with a single plasmonic hotspot~\cite{Ikeda:2013bs,Wasserroth:2018bm}. It was suggested that the confinement of the light fields in space provides the necessary momentum to excite non-vertical optical transitions. This activates the double-resonant Raman process that gives rise to the D-mode in graphene without requiring a real defect for momentum conservation~\cite{Thomsen:2000wf,Reich:2004fb}. As the D-mode in carbon nanotubes has the same origin as in graphene~\cite{Thomsen:2007vc,Reich:2009wo}, we argue that the strong D-mode in the PERS spectra of CNT-B also arises from the localized near-field at the hotspot. This interpretation explains the factor two difference in the experimentally observed $I_D/I_G$ ratios for P$_Y$ and P$_X$. The near-field is less localized for P$_Y$ than it is for P$_X$, resulting in a smaller -- or more likely negligible -- plasmon-induced contribution to the D-mode.

The two-step approach combining DEP and nanomanipulation presented here will find use beyond probing PERS from plasmonic dimers. It will allow to place carbon nanotubes into truly nanoscale gaps ($\leq5\,$nm) of bowtie antennas or plasmonic nanoclusters, which provide the electric field strengths necessary to observe phenomena such as the field gradient effect in PERS~\cite{Ye:2012dv,Kollmann:2014bba,Aikens:2013gm}. The low yield of placing CNT directly in such narrow gaps by DEP makes this technique impractical. Using DEP to carry the nanotubes close to the hotspots followed by nanomanipulation as described here, however, will achieve the desired interface. 

\section{Conclusions}
In conclusion, we suggest directed dielectrophoretic deposition of carbon nanotubes followed by tip-based nanomanipulation to probe plasmon-enhanced Raman scattering from nanoscale plasmonic hotspots. With the tip of an AFM, we pushed a small carbon nanotube bundle $25\,$nm into the gap of an Au nanodimer. Its location at the hotspot was confirmed directly by AFM and indirectly by plasmon-enhanced Raman scattering. We observed a $10^2$ enhancement of the G-mode that was strongly localized in space in combination with an inverted polarization behaviour. Given the ability to characterize the nanotube beforehand, the strong D-mode after nanomanipulation was interpreted as a signature of plasmonic enhancement rather than being caused by structural defects while moving the nanotubes. Our two-fold scheme will allow the reliable placement of CNTs into nanoscale gaps of plasmonic structures and thereby enable experiments previsouly not accessable.\

\section*{Acknowledgements}
The authors acknowledge valuable discussions with S.~Reich and thank A.~Oikonomou for assistance in the sample fabrication. This research was supported by the Engineering and Physical Sciences Research Council (EPSRC) grants EP/K016946/1 (SH, AV) and  EP/G03737X/1 (NC, AV).

\end{document}